\documentclass{egpubl}
\usepackage{egTechReport2019}

\ConferenceSubmission   % uncomment for Conference submission

\usepackage{booktabs} % For formal tables
\usepackage[T1]{fontenc}
\usepackage{dfadobe}  

\usepackage{cite}  % comment out for biblatex with backend=biber
\BibtexOrBiblatex
\electronicVersion
\PrintedOrElectronic
\usepackage{egweblnk} 

\usepackage{adjustbox}
\usepackage{xspace}
\usepackage{relsize}

\usepackage[dvipsnames]{xcolor}
\usepackage{amsmath}
\usepackage{listings}
\usepackage{amssymb}
\usepackage{subcaption}
\usepackage{tabularx}
\usepackage{xcolor}
\usepackage{pgfplots}
\usepackage{tikz}
\usepackage{tabu}

\definecolor{dblue}{rgb}{0,0,0.4}
\definecolor{bblue}{HTML}{4F81BD}
\definecolor{dred}{rgb}{0.4,0,0}
\definecolor{rred}{HTML}{C0504D}
\definecolor{dgreen}{rgb}{0,0.4,0}
\definecolor{ggreen}{HTML}{9BBB59}
\definecolor{dpurple}{rgb}{0.4,0,0.4}
\definecolor{ppurple}{HTML}{9F4C7C}
\definecolor{dyellow}{rgb}{0.4,0.4,0}
\definecolor{yyellow}{rgb}{1,1,0}
\definecolor{dlime}{rgb}{0.1,0.4,0}
\definecolor{llime}{rgb}{0.749,1,0}
\definecolor{dkingblue}{rgb}{0,0.1,0.6}
\definecolor{kkingblue}{rgb}{0,0.2,1}
\definecolor{dgray}{rgb}{0.3,0.3,0.5}
\definecolor{ggray}{rgb}{0.75,0.75,0.85}
\definecolor{dblack}{rgb}{0,0,0}
\definecolor{bblack}{rgb}{0.1,0.1,0.1}

\DeclareCaptionSubType*{figure}

\definecolor{dkgreen}{rgb}{0,0.6,0}
\definecolor{dkblue}{rgb}{0,0,0.6}
\definecolor{gray}{rgb}{0.5,0.5,0.5}
\definecolor{mauve}{rgb}{0.58,0,0.82}

\definecolor{commentgreen}{RGB}{2,112,10}
\definecolor{eminence}{RGB}{108,48,130}
\definecolor{weborange}{RGB}{255,165,0}
\definecolor{frenchplum}{RGB}{129,20,83}

\newcommand{\Test}[1]{\expandafter\hat#1}

\title[Empty Space Skipping Comparison]{Comparing Hierarchical
Data Structures for Sparse Volume Rendering with Empty Space Skipping}
\author[Zellmann]{\parbox{\textwidth}{\centering
    Stefan Zellmann%$^1$
    \thanks{zellmann@uni-koeln.de, Department of Computer Science, University of
    Cologne}\orcid{0000-0003-2880-9090}
    \quad
  }
}

\begin{document}

%\begin{teaserfigure}
%\teaser{ \centering \resizebox{0.98\textwidth}{!}{
    %    \includegraphics[height=5cm]{png/bricks-gear-close.png}
    %    \includegraphics[height=6cm]{png/mag-cropped.png}
    %\fbox{\includegraphics[height=6cm]{png/cloud8K.png}}
    %\fbox{\includegraphics[height=6cm]{png/landinggear.png}}
    %    \includegraphics[height=6cm]{png/landinggear.png}
    %    \includegraphics[height=6cm]{png/deep-water-cropped.png}
    %    \includegraphics[height=5cm]{png/bricks-gear-far.png}
    %    \fbox{\includegraphics[height=6cm]{png/exajet-velocity-0.png}}
    %\fbox{\includegraphics[height=6cm]{png/exajet-velocity-1.png}}
    %    \fbox{\includegraphics[height=6cm]{png/exajet-beauty-cropped.png}}
    %    \fbox{\includegraphics[height=6cm]{png/exajet-wing-cropped.png}}
  %}
  %\resizebox{0.98\textwidth}{!}{
  %  %\includegraphics[height=6cm]{teaser.png}
  %}
%\end{teaserfigure}
%

%\firstsection{Introduction}
\maketitle

\begin{abstract} Empty space skipping can be efficiently implemented with
hierarchical data structures such as \textit{k}-d trees and bounding volume
hierarchies. This paper compares several recently published hierarchical data
structures with regard to construction and rendering performance. The papers
that form our prior work have primarily focused on interactively building the
data structures and only showed that rendering performance is superior to using
simple acceleration data structures such as uniform grids with macro cells. In
the area of surface ray tracing, there exists a trade-off between construction
and rendering performance of hierarchical data structures. In this paper we
present performance comparisons for several empty space skipping data structures
in order to determine if such a trade-off also exists for volume rendering with
uniform data topologies.
\end{abstract}

\section{Introduction}
Ever growing data set sizes resulting from scientific simulations also result in
an increased demand for more efficient rendering algorithms for volumetric data
sets. While strategies like adaptive mesh refinement (AMR) \cite{wang:2019} help
to reduce the memory space required for the data domain, simulations to a large
degree still use uniform grid topologies, so that volume rendering with
structured grids is still an active area of research. Structured grids also
result from medical imaging techniques such as computer tomography (CT) or
magnetic resonance imaging (MRI). Data sets from the clinical context in
practice often have low resolution (for such reasons as to avoid unnecessarily
exposing patients to radiation), but on the other hand often consist of multiple
modalities like fiber tracks or blood vessels that were derived or segmented
from the original CT or MRI data and manifest as separate volume channels. Those
derived volume channels are often sparse in nature.

In all the described settings, it is beneficial if the user can explore the 3-d
data set by means of interaction. A part of this interaction is adjusting the
color and alpha transfer function that maps scalar input from the volume field
to (RGB) color and opacity. When the alpha transfer function changes, so does
the overall amount of empty space as well as the spatial arrangement of
non-empty voxels.

A number of recent publications from our research group
\cite{zellmann:2018,zellmann:2019,zellmann:2019b,zellmann:2019c} has therefore
concentrated on the construction performance of spatial indices for direct
volume rendering and proposed several data structures that can - depending on
the size of the data set - be built at interactive rates. As the construction
times reported in the papers vary significantly, in this paper we present a
thorough performance comparison of the techniques.

The construction algorithm from~\cite{zellmann:2019} is based on the linear
bounding volume hierarchy (LBVH) algorithm~\cite{lauterbach:2009} and shows
significantly improved construction rates compared to the other more involved
data structures.  In the field of real-time ray tracing with surface geometry,
the LBVH data structure is generally known for its inferior culling properties
but superior construction performance. The data structures from
\cite{zellmann:2018} and \cite{zellmann:2019b} especially employ very intricate
construction schemes that use a cost function comparable to the surface area
heuristic (SAH)~\cite{wald:2007} known from surface ray tracing. In that field,
it is generally accepted that the quality obtained with SAH is usually higher
than that of LBVH, which uses the middle split heuristic. The construction time
for hierarchies built with SAH however is generally higher than the construction
time required when using a simpler heuristic.

While the various papers from our prior work individually proved the
effectiveness of the respective data structures in general, they only compared
them with na\"ive ray marching without empty space skipping, or with simple data
structures like structured grids with macro cells. This paper tries to fill the
gap in this respect and provides a comparison of the various data structures
against each other. This is done by testing the data structures using different
combinations of data sets and transfer functions.

This paper shall be understood as an addendum to our prior work: the benchmark
results that constitute the \textbf{main contribution} of this paper are meant
to serve as a guideline and to give further insight into when to use which of
the data structures we proposed in our recent papers on empty space skipping.

\section{Related Work}
Interactivity is an important property for scientific visualizations and can be
aided in several ways, such as via remote rendering
\cite{zellmann:2012,shi:2015}, level-of-detail techniques~\cite{leven:2002} or
adaptive sampling of the data domain~\cite{morrical:2019}. Empty space skipping
is one of the traditional optimization strategies for direct volume rendering
(DVR) algorithms and is often implemented using hierarchical data
structures~\cite{li:2003,hadwiger:2012,liu:2013}. A general overview on empty
space skipping and other optimization techniques can e.g.\ be found in the
state-of-the-art report by Beyer et al.~\cite{beyer:2014}. Notable systems or
system approaches are those by Hadwiger et al.
\cite{hadwiger:2012,hadwiger:2018} or the grid-based solution that can be found
in OSPRay~\cite{wald:2017}. The system by Hadwiger et al. and the OSPRay
implementation re\-present two different approaches to fundamentally solve the
same problem; Hadwiger's system is based on rasterization and performs volume
integration in a shader, while OSPRay implements volume rendering as a genuine
ray tracing program. While OSPRay does not make use of hardware acceleration,
researchers have recently shown the effectiveness of ray tracing hardware for
volume rendering with ray casting and empty space skipping
\cite{ganter:2019,morrical:2019}.

Hardware accelerated ray tracing uses acceleration data structures that are
described by a whole corpus of research papers that is primarily focused on
surface ray tracing, e.g.\
\cite{wald:2006,wald:2007,karras:2012,gu:2013,meister:2018}. While some
researchers have used acceleration data structures to implement volume rendering
algorithms in the past~\cite{zellmann:2017}, with the advent of hardware
accelerated ray tracing on modern GPUs it is likely that the two research
branches that are concerned with surface ray tracing on the one hand, and with
volume ray casting on the other hand, are going to converge in the future.

While interactive hierarchy rebuilds have gained a significant amount of
attention from the real-time ray tracing community in the past
\cite{lauterbach:2009,pantaleoni:2010,meister:2018}, interest from scientific
visualization researchers has only recently started to focus on this topic.
Our group has recently proposed a number of empty space skipping data structures
\cite{zellmann:2018,zellmann:2019,zellmann:2019b,zellmann:2019c} that are
primarily aimed at fast spatial index reconstruction and are explained in detail
in the following section. While our approach is focused on rebuilding the
hierarchy whenever the data or the transfer function changes, another approach
is to reuse or adapt an already existing acceleration data structure. This can
e.g.\ be achieved by means of \texttt{min-max} range queries
\cite{wald:2007b,knoll:2011,wald:2019}. An alternative approach to update an
existing data structure is the one by Schneider et al.~\cite{schneider:2017} who
use Fenwick trees, which bare similarities to the summed area tables that our
techniques use as auxiliary data structures.

\section{Construction Algorithms for Empty Space Skipping Hierarchies}
In this section we briefly summarize the various construction algorithms from
our recent papers. While the linear bounding volume hierarchy algorithm is
solely targeted towards GPUs, the \textit{k}-d construction algorithm we have
targeted towards both multi-core and GPU systems. The hybrid grid construction
algorithm is based on the multi-core CPU variant of the \textit{k}-d tree
construction algorithm but could generally also be implemented on the basis of
the GPU variant.

\subsection{Linear bounding volume hierarchies}
The LBVH algorithm was initially introduces by Lauterbach et
al.~\cite{lauterbach:2009}. Our implementation~\cite{zellmann:2019} adapts this
algorithm to volume rendering with uniform grids. The implementation uses NVIDIA
CUDA. We first decompose the volume into bricks of size $8^3$ and then run a
CUDA kernel where each thread is responsible for one voxel. Each thread
determines if the voxel is visible w.r.t.\ the current transfer function. The
threads responsible for one brick then \emph{vote} if the whole brick is visible
by atomically updating a flag in on-device shared memory. This per-voxel
operation is followed by a number of operations with $O(n)$ work
complexity---where $n$ is the number of bricks---each of which is carried out in
a single CUDA kernel. We first perform compaction since we are only interested
in the non-empty bricks.  Then we assign 30-bit 3-d Morton codes to the
non-empty bricks that we use to sort the bricks using a parallel $O(n)$ GPU
algorithm from the Thrust library~\cite{bell:2011}. The Morton order implicitly
defines a hierarchy over the bricks that just spatially splits the respective
child nodes in the middle. The split positions can be read from the bit codes of
the Morton indices and can be efficiently found using Karras'
algorithm~\cite{karras:2012}. A final CUDA kernel traverses the hierarchy from
each leaf node up to the root node and assembles the respective axis-aligned
bounding box of each inner node encountered along that path.

\subsection{\textit{k}-d trees}
The \textit{k}-d tree construction algorithms from~\cite{zellmann:2018}
and~\cite{zellmann:2019b} are loosely based on original work by Vidal et
al.~\cite{vidal:2008} which employs a summed area table to quickly determine the
occupancy (i.e.\ the number of non-empty voxels) inside a volumetric region
bound by an axis-aligned bounding box. We first introduce the multi-core
parallel variant of the construction algorithm that will produce the exact same
results as the algorithm by Vidal et al. (apart from certain parameters such as
halting criteria etc. that we might set differently) and then briefly describe
the adaptation of this algorithm to the GPU.

\subsubsection{Multi-core construction}
With the x64 multi-core CPU implementation that was first presented in
\cite{zellmann:2018} and later refined in~\cite{zellmann:2019b}, the volume is
first decomposed into bricks of size $32^3$. A preclassified version where each
voxel is only associated with a flag telling whether it is empty or not is
derived from that whenever the transfer function changes. We then build a
three-dimensional summed area table (a ``summed volume table'', SVT) over the
binary preclassification for each brick. By choosing a brick size of $32^3$, an
SVT will fit exactly into the L1 cache of an x64 CPU. The respective SVTs for
each brick are built in parallel.

This SVT construction phase is followed by a second algorithmic phase where a
\textit{k}-d tree is built in a top-down fashion. We first determine a tight
bounding box for the root node by shrinking the axis-aligned bounding box of the
whole volume to tightly fit the non-empty voxels according to the current
transfer function. This can be done by querying the SVTs to determine the
occupancy inside the boxes. Occupancy queries are performed in parallel for each
SVT that the bounding box overlaps. Therefore, local bounding boxes are computed
in parallel per SVT and the result is later combined to form the overall
bounding box. The shrinking procedure just compares the occupancy inside smaller
boxes to the occupancy of the parent bounding box; if the occupancy is the same,
shrinking was valid. A cost function is used to determine an optimal splitting
plane by using sweeping and by inspecting the volume of the (tight) bounding box
to the left and the right of the candidate planes. Certain halting criteria
favor either shallow or deep trees---both of which we evaluate in
Section~\ref{sec:results}---and can be set as parameters to the construction
algorithm. The targeted size (height; number of nodes) of the hierarchy one can
expect to influence both construction time and rendering performance.

\subsubsection{Construction on GPUs}
\label{sec:gpu}
The CPU construction algorithm---would it be ported without modifications---is
not well suited for execution on GPUs, at least not if the whole algorithm would
be carried out in a single GPU kernel. In~\cite{zellmann:2019b} we therefore
proposed an adapted version of the algorithm that performs the plane sweeping
and top-down construction phase of the algorithm on the CPU, while the procedure
that finds tight bounding boxes around non-empty voxels is off\-loaded to the GPU.

The plane sweeping procedure thus involves starting two GPU kernels for both
the left and the right half-space and for each plane tested against. To reduce
the number of kernel calls, we use a binning approach. Binning has another
crucial advantage: strategically aligning the bin boundaries on the same raster
imposed by the SVTs, we will never consider plane candidates that would split an
SVT into two halves and can thus just precompute the tight bounding box inside
that macro cell once when the transfer function changes (as opposed to each time
that we consider another plane candidate). We do this by initially computing
SVTs that are however immediately discarded as soon as the local bounding box
has been determined. In contrast to the CPU, where we optimized for L1 memory,
on the GPU we explicitly perform the computations in shared memory, so that a
macro cell will have a size of $8^3$.

We order the resulting bounding boxes on a z-order Morton curve. In order to
determine a tight bounding box for a spatial region spanning multiple macro
cells during splitting, we perform a parallel reduction on the GPU using Thrust.
Since this is a 1-d operation, we need to check if the macro cells we consider
are actually inside the volumetric region of interest. To minimize the number of
macro cells to test, we first conservatively determine the first and last cell
in the list that will definitely fall inside the region of interest by using the
Morton code order of the list. Although the algorithm is per se not very well
suited for GPUs due to the top-down recursion, we still achieve good GPU
utilization during splitting as we perform the reduction twice for each
candidate plane, and also because of the smaller size and thus larger amounts of
the macro cells to reduce over.

\subsection{Hybrid grids}
Our benchmarks suggested that shallow \textit{k}-d trees built with the original
halting criteria proposed by Vidal et al. would result in effective space
skipping data structures, but on contemporary hardware, deeper trees with a leaf
node size of approximately $8^3$ would perform even better. The approach by
Vidal et al. was originally intended to be used in a way where macro blocks
would be rendered using an outer loop, which calls a volume rendering shader per
block. In~\cite{zellmann:2019c} we evaluated the outer loop approach against
full ray traversal on the GPU; we found that full ray traversal with relatively
deep trees would outperform the outer loop approach during rendering, but also
that the construction times for deep trees was significantly higher than that
for shallow trees.

We therefore proposed an alternative rendering strategy, which would combine a
shallow \textit{k}-d tree with only relatively few leaves with a global grid of
macro cells. As the \textit{k}-d tree construction algorithm effectively uses
macro cells---each SVT can be thought of as a macro cell that stores occupancy
information about its volumetric region---deriving a grid from that and sending
it over to the GPU is straightforward and comes at no recognizable storage
overhead. Instead of the original size of $32^3$ that the construction algorithm
uses, we found macro cells of size $16^3$ in general to perform better and thus
resample the grid to that size, which can still be done in constant time using
the SVTs. We adapted the ray marching algorithm to perform full traversal per
ray until we find a \textit{k}-d tree leaf node, and inside that one use the
global grid to skip over empty space with finer granularity. Our benchmarks
suggest that the data structure is helpful in certain cases, specifically when
the data sets are large (i.e.\ near the amount of available texture memory), or
when empty space manifests as wholes inside the volume. Construction performance
however was literally the same as that for shallow \textit{k}-d trees.

\section{Performance Comparisons}
\label{sec:results}
\begin{figure*}
  \setlength\tabcolsep{.5ex}
  \centering
  \begin{tabular}{cccc}
      \includegraphics[width=.24\textwidth]{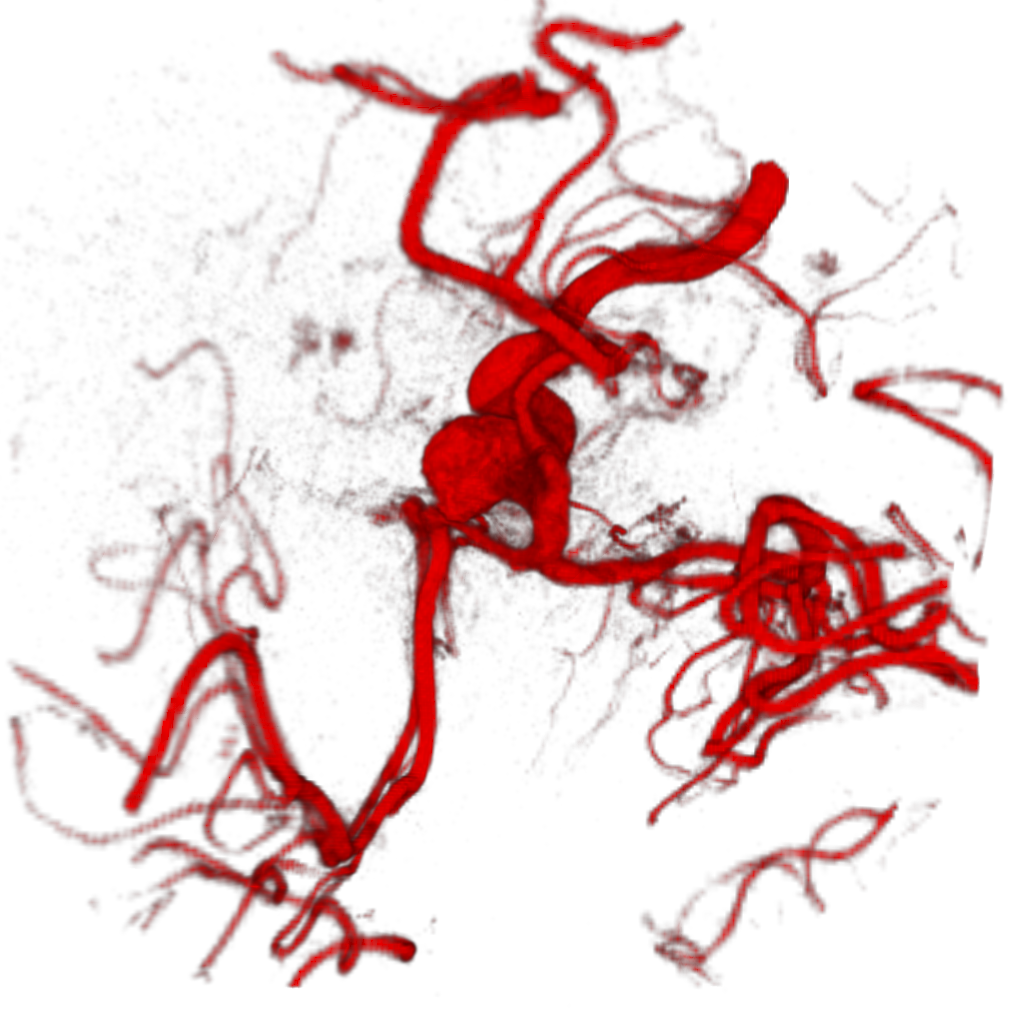}&
      \includegraphics[width=.24\textwidth]{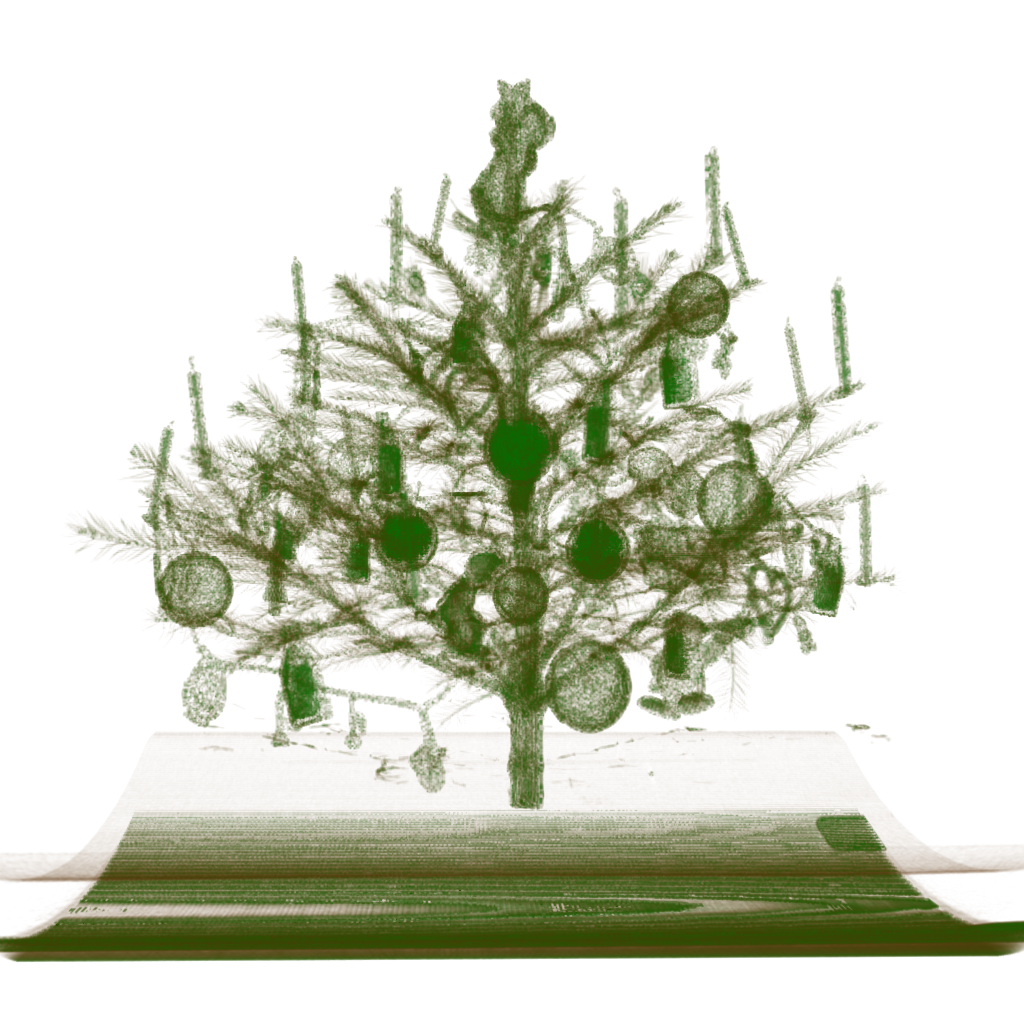}&
      \includegraphics[width=.24\textwidth]{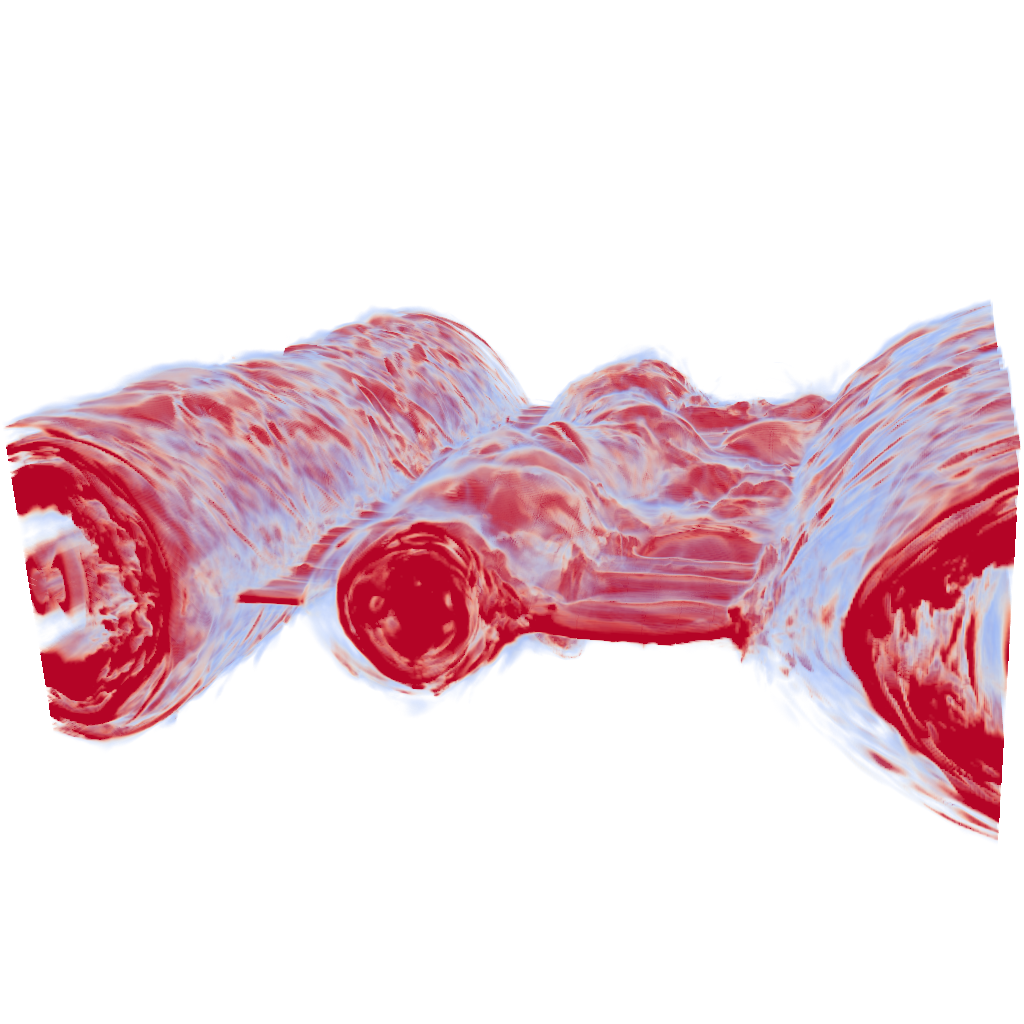}&
      \includegraphics[width=.24\textwidth]{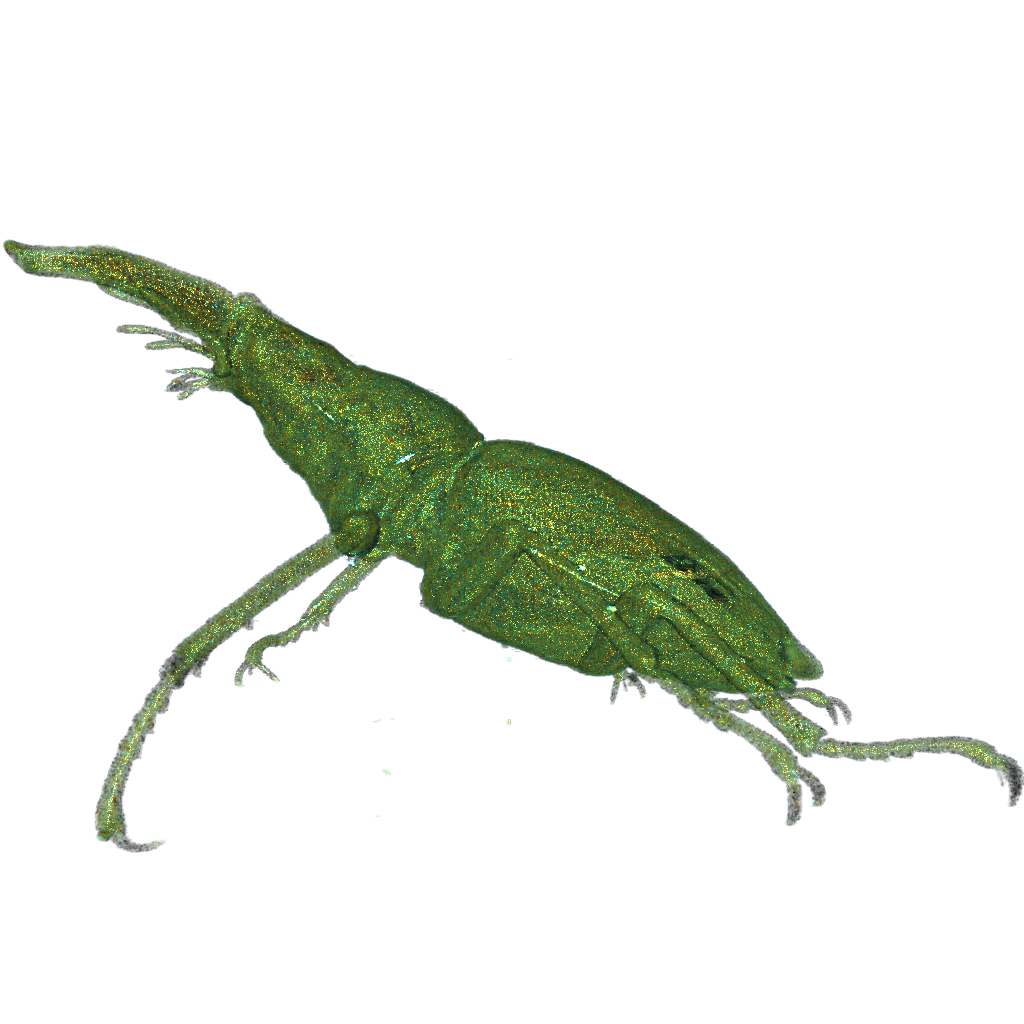}
      \\
      \small Aneurism
      &
      \small Xmas Tree
      &
      \small Magnetic Reconnection Simulation
      &
      \small Stag Beetle
      \\
      \small $256^3$ Voxels
      &      
      \small $512 \times 499 \times 512$ Voxels
      &      
      \small $512^3$ Voxels
      &      
      \small $832 \times 832 \times 494$ Voxels
      \\
      \small Occupancy: $1.01~\%$
      &      
      \small Occupancy: $2.90~\%$
      &      
      \small Occupancy: $16.6~\%$
      &      
      \small Occupancy: $4.04~\%$
      \\
      \includegraphics[width=.24\textwidth]{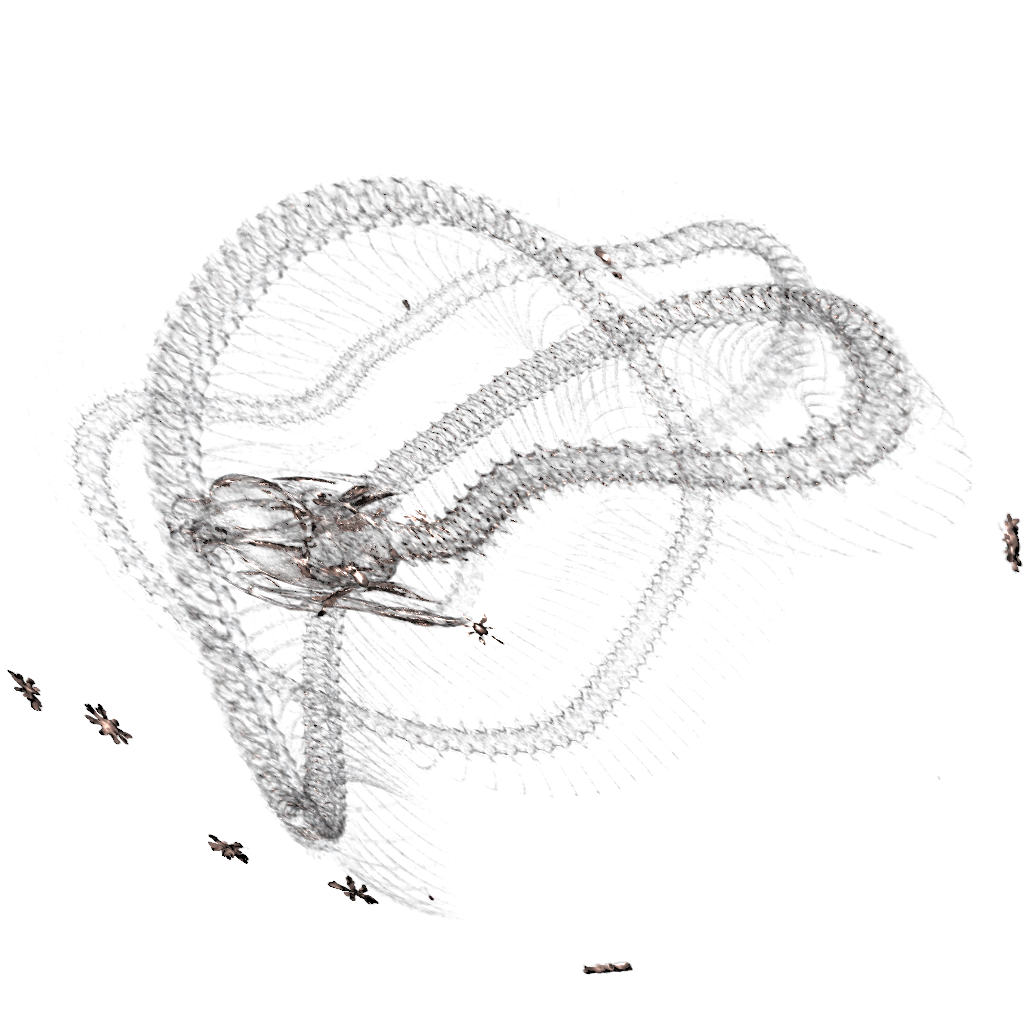}&
      \includegraphics[width=.24\textwidth]{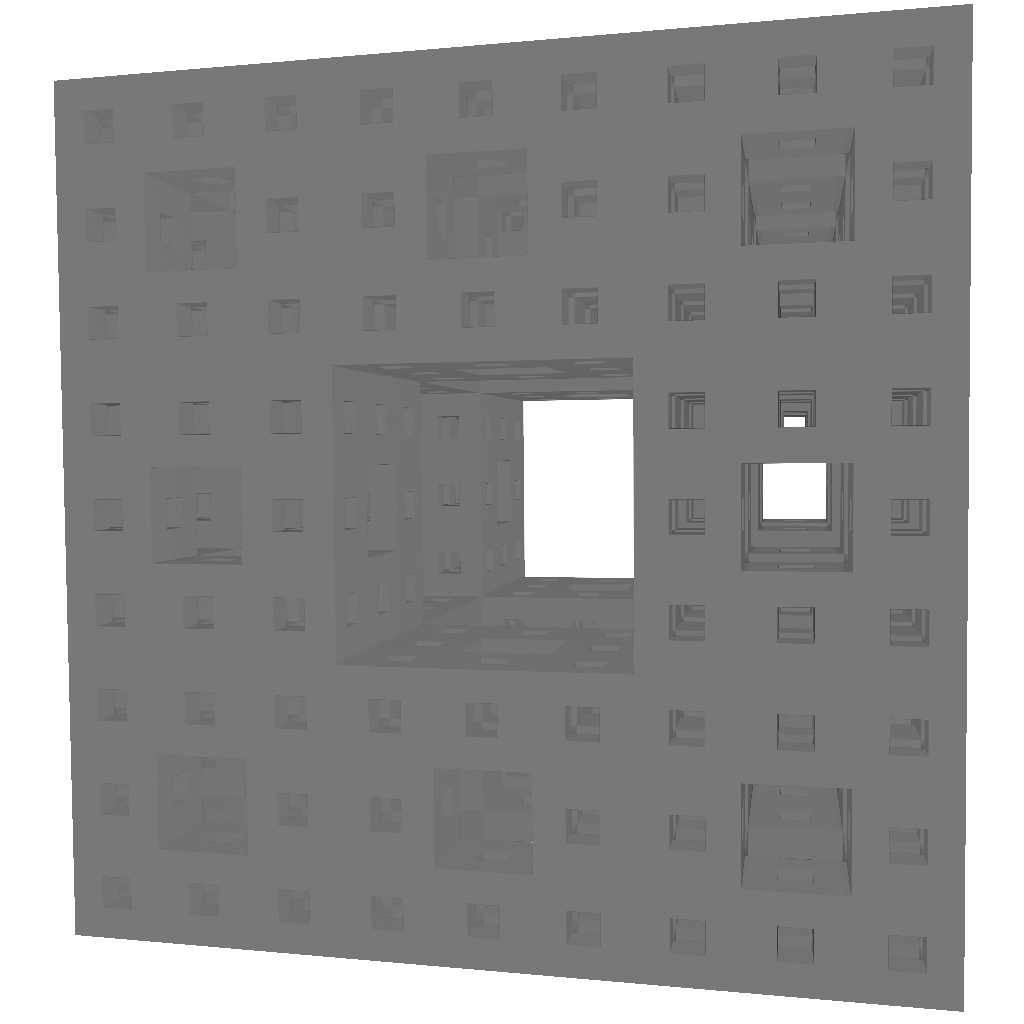}&
      \includegraphics[width=.24\textwidth]{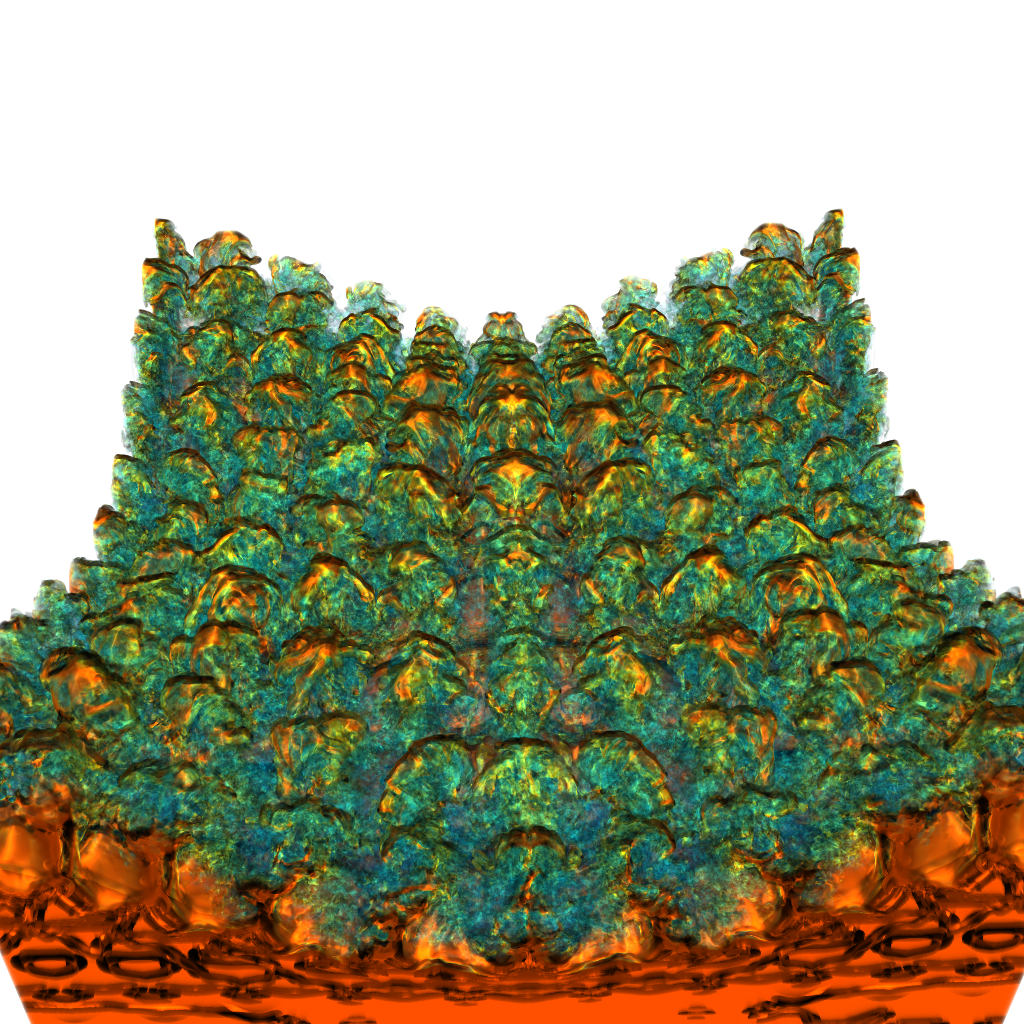}&
      \includegraphics[width=.24\textwidth]{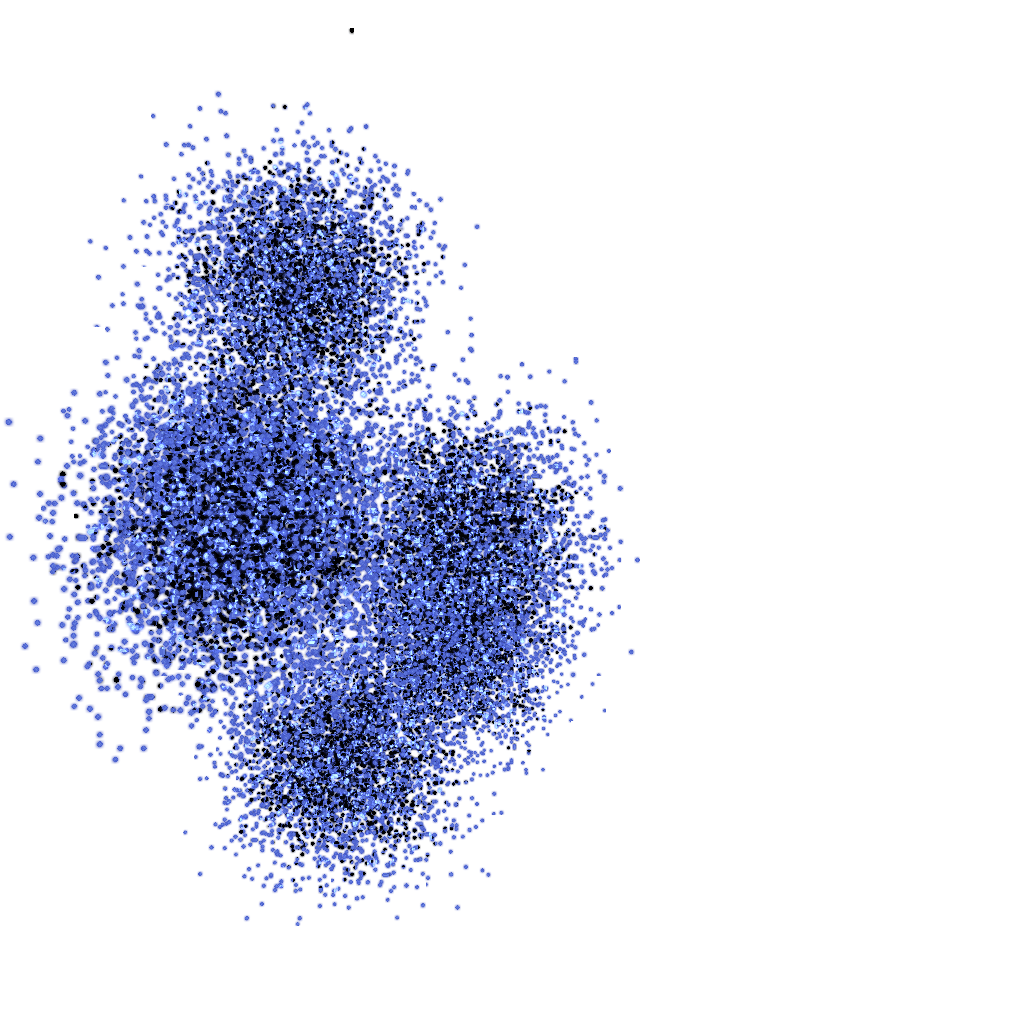}
      \\
      \small Kingsnake
      &
      \small Menger Sponge
      &
      \small Richtmyer-Meshkov Instability
      &
      \small N-Body
      \\
      \small $1024 \times 1024 \times 795$ Voxels
      &      
      \small $1024^3$ Voxels
      &      
      \small $2048 \times 2048 \times 1920$ Voxels
      &      
      \small $256^3$~/~$512^3$~/~$1024^3$~/~$2048^3$ Voxels
      \\
      \small Occupancy: $0.39~\%$
      &      
      \small Occupancy: $40.7~\%$
      &      
      \small Occupancy: $50.6~\%$
      &      
      \small Occupancy: $0.15~\%$
      \\
  \end{tabular}
  %\vspace{1mm}
  \caption{\label{fig:models}%
    Data sets with spatial dimensions and occupancy (percentage of voxels that
    are visible given a certain transfer function) used to evaluate our method.
    We pick a number of different settings with transfer functions that favor
    different types of spatial arrangements of the visible voxels.
    %\vspace*{-5mm}
  }
%
%    \iw{placeholder for now - use all square images, or all 1024**2; as pretty as we can; ideally all easily reproduible (for
%      results table) 
      %
%    \nvm{Probably replace the last LANL impact picture with something more interesting}
\end{figure*}
%--- max leaf size figure -----------------------------------------------------
\begin{figure*}
  \setlength\tabcolsep{.5ex}
  \centering
  \begin{tabular}{ccc}
      \includegraphics[width=.33\textwidth]{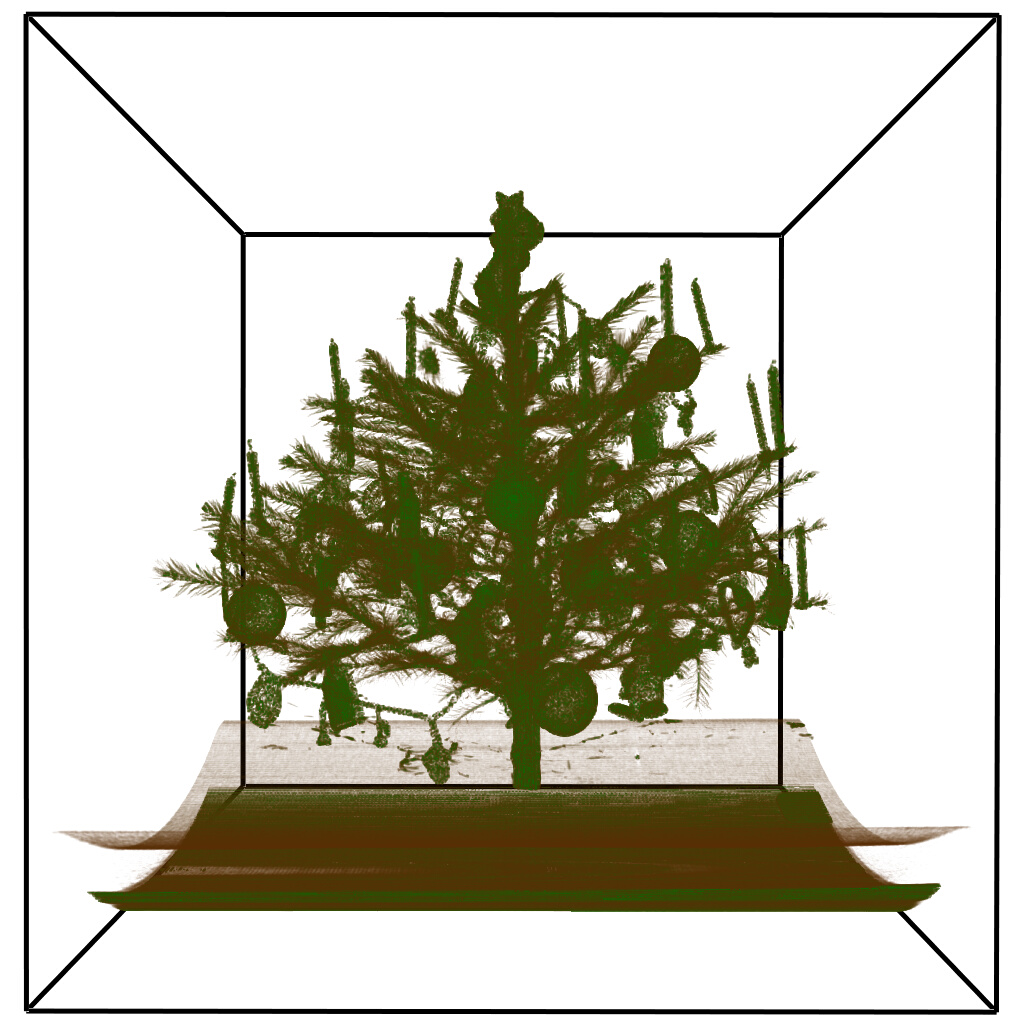}&
      \includegraphics[width=.33\textwidth]{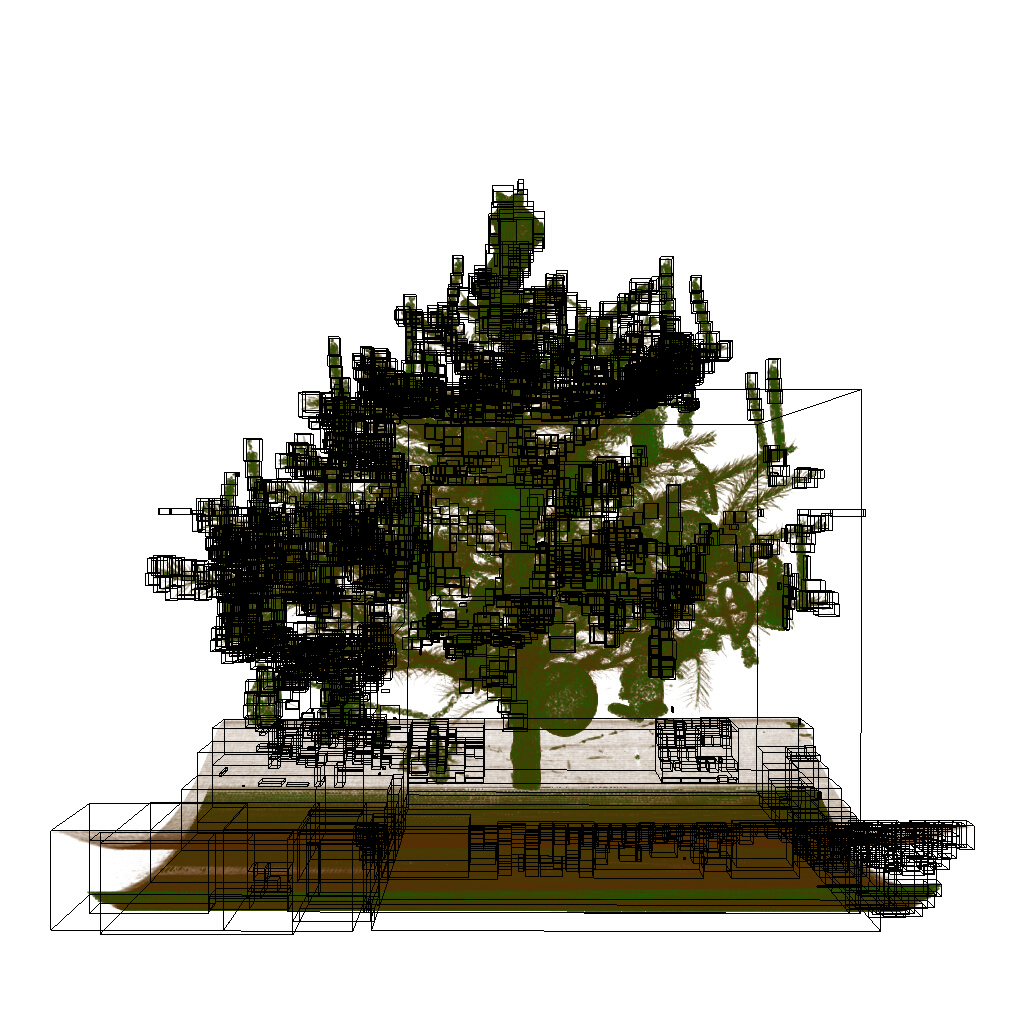}&
      \includegraphics[width=.33\textwidth]{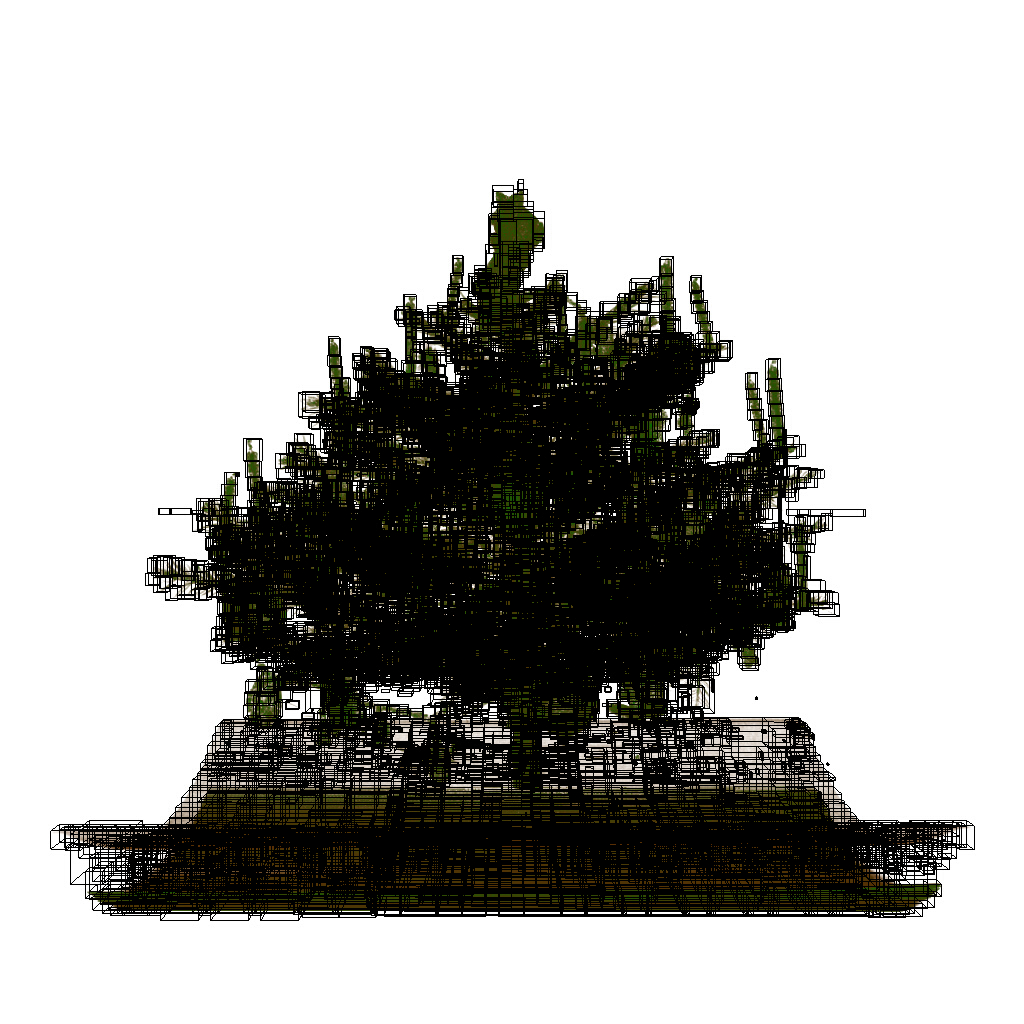}\\
      \includegraphics[width=.33\textwidth]{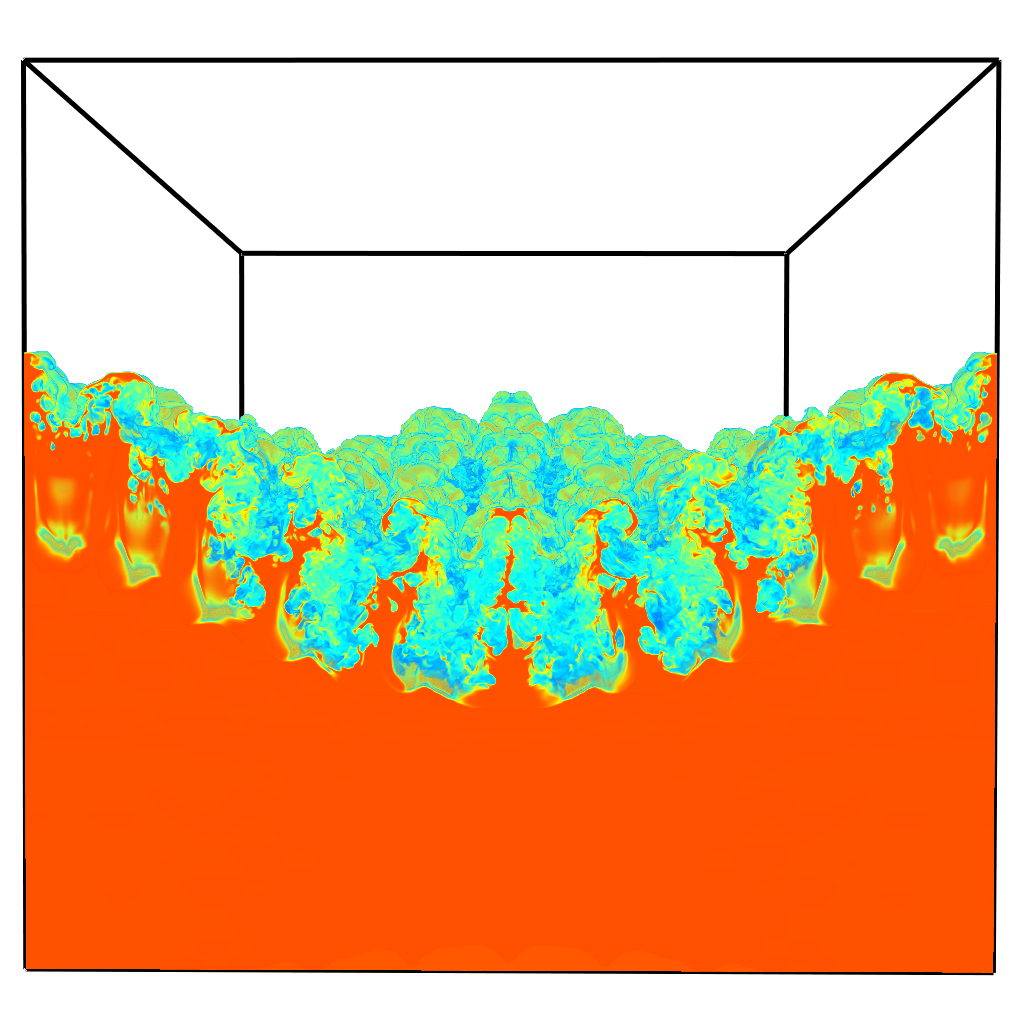}&
      \includegraphics[width=.33\textwidth]{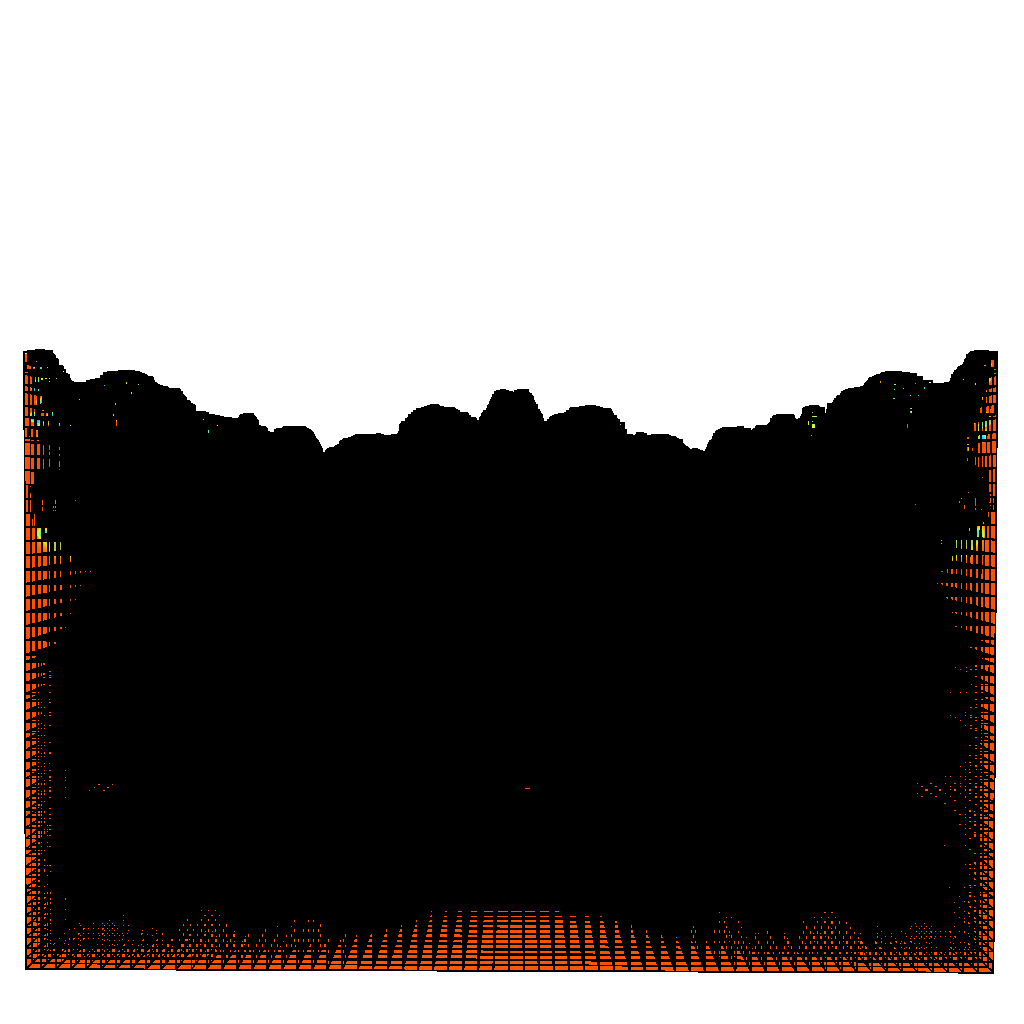}&
      \includegraphics[width=.33\textwidth]{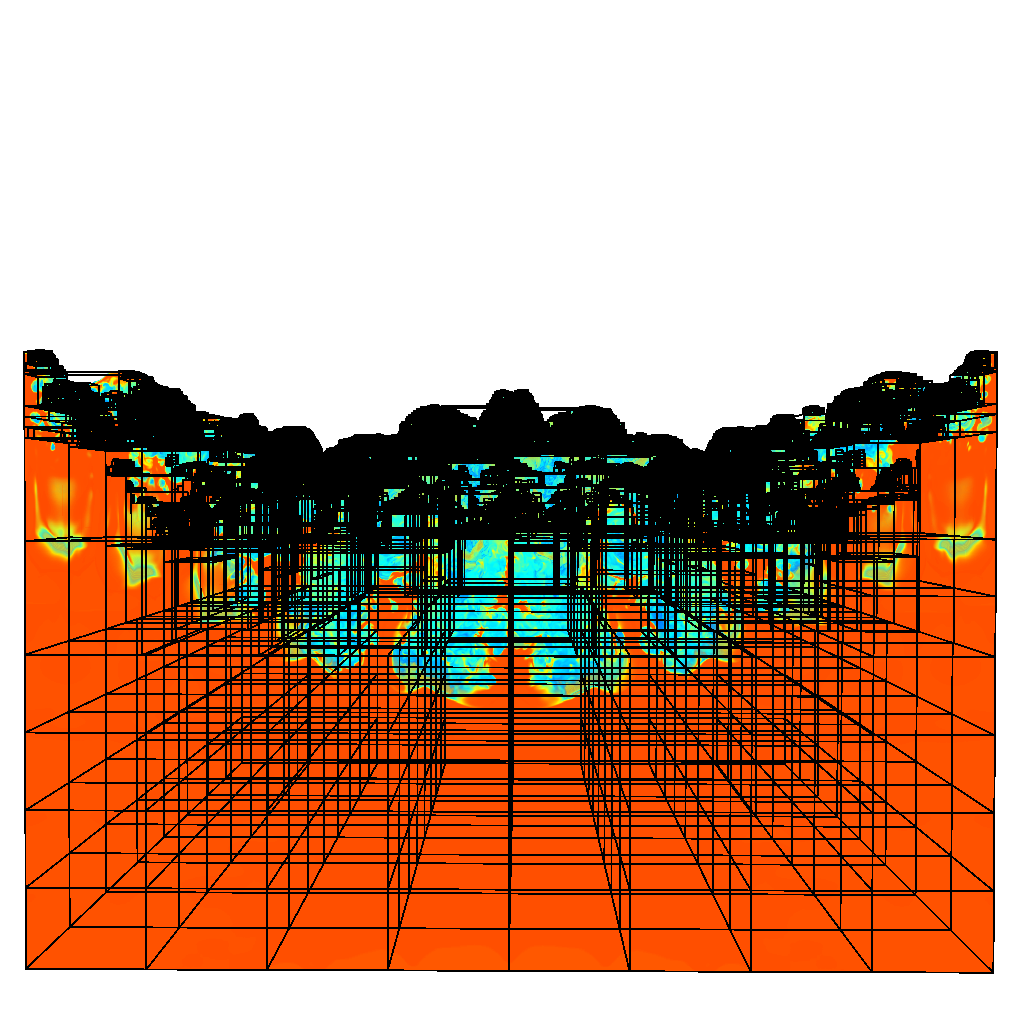}
  \end{tabular}
  \caption{\label{fig:splitting}%
    Enforcing a maximum leaf size to avoid unfavorable local optima.  Top row,
    left: Xmas tree data set with bounding box overlay. Building a deep
    \textit{k}-d tree for this data set and transfer function combination will
    cause the splitting heuristic to accept a local optimum resulting in a
    single very large leaf node (middle). Enforcing a maximum leaf node size can
    mitigate this issue (right). If the volume consists of regions that are
    non-empty (bottom row, left), enforcing a maximum leaf size can however have
    the undesirable effect of the hierarchy growing exceptionally deep and wide
    and splitting regions containing no empty space (middle). A sensible choice
    for the maximum leaf size depends on the data set and the transfer function
    (right). We consider investigating this issue and potentially finding better
    solutions interesting future work.
    }
\end{figure*}
%--- table --------------------------------------------------------------------
\begin{table*}[tb]
    %rendering performance for na\"ive ray marching without space skipping. In parentheses we report
    %the rendering times for a transfer function that does not assign empty space at all. Data structures that depend
    %on \emph{min-max} range queries incur an extra constant overhead of $0.042$ seconds to build an IRT for a
    %transfer function array with $1024$ entries.}
	\centering%
    \footnotesize
    \setlength\tabcolsep{4pt}
  \begin{tabu}{l|r|r|r|r|r|r|r|r|r|r|r|r|r|r|}
    \toprule

Data Set &
\multicolumn{2}{c|}{LBVH} &
\multicolumn{2}{c|}{CPU Shallow} &
\multicolumn{2}{c|}{CPU, MLS=32} &
\multicolumn{2}{c|}{CPU, MLS=128} &
\multicolumn{2}{c|}{GPU Shallow} &
\multicolumn{2}{c|}{GPU, MLS=32} &
\multicolumn{2}{c|}{GPU, MLS=128} \\

\midrule
~ &
\#~nodes & height &
\#~nodes & height & 
\#~nodes & height &
\#~nodes & height & 
\#~nodes & height &
\#~nodes & height & 
\#~nodes & height \\

\midrule

Aneurism        & 14,081    & 16 &  39& 12	  & 503	    & 32 &	3,003	& 26 &
23  & 7  & 8,895	& 29& 3,571   & 22   \\
Xmas Tree       & 85,089    & 19 &  33& 11	  & 21,041  & 38 &	13,453	& 37 &
23  & 7  & 35,677	& 38& 16,753  & 31  \\  
Magnetic        & 110,253   & 18 &  51& 14	  & 12,329  & 37 &	4,265	& 34 &
29  & 6  & 20,663	& 31& 7,261   & 26  \\
Stag Beetle     & 106,923   & 21 &  19& 7	  & 14,229  & 31 &	8,615	& 29 &
21  & 7  & 23,559	& 36& 13,747  & 28  \\
Kingsnake       & 92,047    & 22 &  31& 14	  & 33,035  & 46 &	30,359	& 46 &
17  & 7  & 40,619	& 39& 34,087  & 37  \\
Menger Sponge   & 2,002,859 & 22 &  1 & 1	  & 115,335 & 22 &	1,399	& 13 & 1
& 1  & 181389	& 41& 7,269   & 28  \\
Richtmeyr       & 18,191,777& 25 &  3 & 2	  & 764,809 & 40 &	99,169	& 35 & 1   & 1  & 2,287,401	& 48& 166,745 & 36  \\
N-Body $256^3$  & 8,753     & 16 &  21& 8	  & 2,917   & 26 &	2,095	& 24 &
15  & 8  & 5,021	& 30& 3,039   & 22  \\
N-Body $512^3$  & 41,699    & 19 &  21& 8	  & 13,427  & 34 &	9,367	& 34 &
17  & 9  & 22,823	& 35& 10,687  & 31  \\
N-Body $1024^3$ & 127,572   & 22 &  25& 10	  & 39,287  & 59 &	28,991	& 53 &
17  & 9  & 58,189	& 41& 30,161  & 36  \\
N-Body $2048^3$ & 361,633   & 25 &  23& 10	  & 66,539  & 72 &	64,047	& 71 &
19  & 10 & 94,485	& 42& 80,943  & 40  \\

    \bottomrule
  \end{tabu}
  \caption{\label{tab:statistics}
    Statistics for the various data sets we use for the evaluation. We report
    the number of nodes and the height of the respective tree structures:
    Linear bounding volume hierarchies with leaves of size $8^3$.
    Shallow \textit{k}-d trees where a leaf is created when the volume of the
    node gets below $10~\%$ of the root node's volume during splitting;
    \textit{k-d} trees with leaf volume less or equal $8^3$ and a maximum leaf
    node size of $32^3$;
    \textit{k-d} trees with leaf volume less or equal $8^3$ and a maximum leaf
    node size of $128^3$; the \textit{k}-d trees are built with different
    algorithms depending on whether they are built on the CPU or the GPU;
    on the GPU we use four bins to determine candidate split planes
    (cf. Section~\ref{sec:gpu})
    }
\end{table*}
\begin{figure*}
\includegraphics[width=\textwidth]{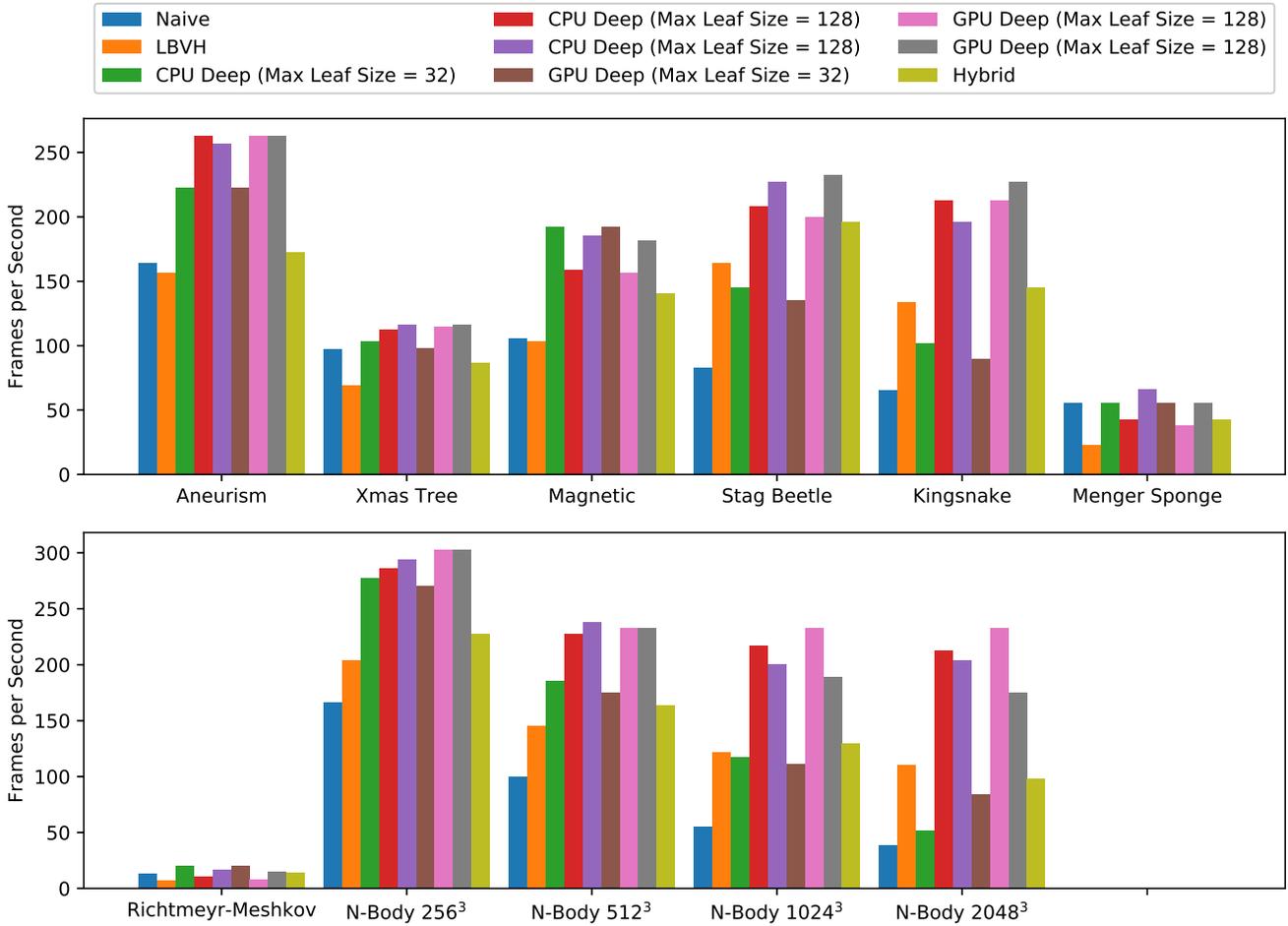}
\caption{\label{fig:fps}
Rendering performance obtained with our benchmarks. We render rotating
orthographic views with a $1024 \times 1024$ pixel viewport. For our tests, we
disable gradient shading and early-ray termination. For the absolute timing
results cf. \autoref{tab:results}.}
\end{figure*}
\begin{figure*}
\includegraphics[width=\textwidth]{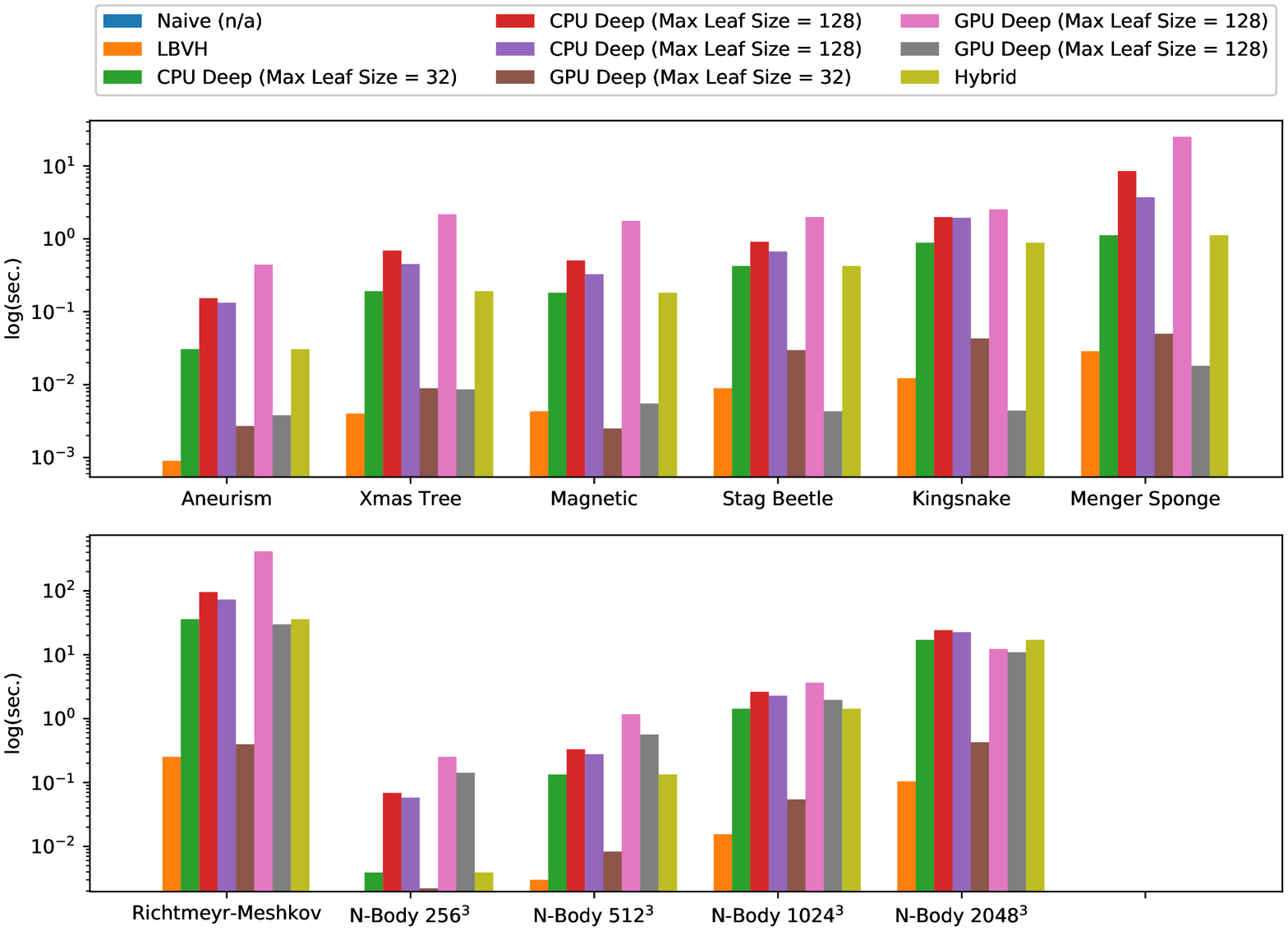}
\caption{\label{fig:construction}
Build rates for the various data sets and hierarchy construction algorithms.
Note the logarithmic scale of the y-axis. For the absolute timing results cf.
\autoref{tab:results}.}
\end{figure*}
The main focus of this paper is a thorough comparison of the various data
structures w.r.t.\ construction and rendering performance. To achieve this, we
integrated them into the Virvo volume rendering library \cite{schulze:2001} and
use the Visionaray library~\cite{zellmann:2017b} to implement the low level ray
tracing algorithms like \textit{k}-d tree or BVH traversal.

\subsection{Test setup}
Our test system consists of an eight-core Intel Xeon Gold 5122 CPU system with
3.60~GHz clock frequency and an NVIDIA Titan-V graphics card with 12~GB GDDR
video memory. The system is equipped with 96~GB DDR memory. For the \textit{k}-d
tree and hybrid grid construction algorithms, we deactivate simultaneous
multithreading (marketed by Intel under the name ``hyperthreading'') and assign
a single core to each thread using the \texttt{numactl} tool that comes with the
Linux distribution installed on our test system.

For the evaluation we use the data sets and transfer function settings
from~\autoref{fig:models}. Note that~\autoref{fig:models} does not depict the
exact view points we used for the evaluation. Rather than that, we use an
orthographic camera and a view point that is zoomed in so that the volume fills
the whole rendering window. We use a CUDA-based renderer with the absorption
plus emission model and post-classification transfer function lookups; for our
tests we deactivate gradient shading and early-ray termination. For the
benchmarks we use a rotating camera animation to account for and average out
effects like unfavorable offsets and strides when accessing 3-d textures from
certain viewing angles. We render images into viewports of size $1024 \times
1024$ pixels. In addition to the rendering performance, we also evaluate the
tree construction performance.

\subsection{Data sets}
We use a variety of different data set and transfer function combinations that
are depicted in~\autoref{fig:models}. We deliberately choose combinations with a
varying number of non-empty voxels. Since all algorithms are in-core, the
largest data sets we test with have a size of $2048^3$ voxels, which amounts to
8~GB when voxels are stored with 8 bit precision. The Richtmeyr-Meshkov
instability specifically has two prominent large regions that are either empty,
or not empty but totally homogeneous. We expect the different data structures to
adapt differently to the large amount of empty space. The Menger Sponge data set
we consider a hard case for most empty space skipping data structures as the
empty space is contained inside the volume, a spatial arrangement which is known
to be particularly ill suited for typical \textit{k}-d tree builders.

\subsection{Parameters}
We consider three different types of \textit{k}-d trees: We build shallow trees
where we only split a node if its volume is above 10~\% of the root node's
volume (that is the original halting criterion proposed by Vidal et
al.~\cite{vidal:2008}). This setup will create large volume chunks which will
potentially contain lots of empty space. The two remaining setups employ a
halting criterion where a node is split whenever its volume is above $8^3$ and
will result in deeper trees.

The greedy heuristic will in certain cases accept a local optimum and thus stop
the recursion due to the cost function (cf.~\autoref{fig:splitting}), even
though the resulting leaf's extent is quite large and still bounds a substantial
amount of empty space. That in particular happens when empty space is contained
inside the volume and was already pointed out by Vidal et al.~\cite{vidal:2008}
This generally undesirable behavior can be mitigated by enforcing a split when
the leaf node would otherwise exceed a certain size. We perform benchmarks for
configurations where we enforce a maximum leaf node size of either $32^3$ or
$128^3$; when the cost function reports a leaf that exceeds the maximum size, we
just split it in two at the middle position of the axis where the spatial extent
is largest and continue the recursion. Note that this procedure will also
enforce splitting large areas of homogeneous or non-empty space (like it is
present in the Richtmeyr-Meshkov and Menger Sponge data sets). The increase in
tree depth due to this behavior may be potentially undesirable, which is why we
include results for two different maximum leaf node sizes. See
\autoref{tab:statistics} for some statistics regarding the tree height and
number of nodes when building the data structures for our test data sets.

Like in~\cite{zellmann:2019b}, the GPU construction algorithm uses binning,
while the CPU construction algorithm does not. We set the number of bins to $4$.
See~\cite{zellmann:2019b} for a thorough evaluation of the influence of binning
on construction and rendering performance. We report the rendering performance
in~\autoref{fig:fps} and the construction performance in
\autoref{fig:construction}. The absolute numbers plotted in the two figures can
also be found in~\autoref{tab:results}.

\subsection{Discussion and Limitations}
Our benchmarks suggest that construction with the LBVH algorithm will be very
fast, but as this algorithm makes the least informed decision as of the position
where to perform the split will generally result in inferior spatial indices.
Out of the several data structures that use SVTs to find tight bounding boxes,
we see a clear correlation between the depth of the resulting tree and the
construction performance, but also observe that deeper trees will generally be
superior to more shallow ones.

The problem with the algorithms accepting local optima can only partially be
mitigated using a maximum leaf size. In particular, this strategy might fail
when the data set is huge and extra splits of homogeneous space that would
otherwise have been bound by a single leaf cause the resulting spatial index to
be deeper and contain far more leaf nodes than without using this strategy. This
behavior can be seen in \autoref{fig:splitting} for the
Richtmeyr-Meshkov data set, where the homogeneous space to the bottom of the
data set is split excessively and to no avail. In the future, we intend to
investigate alternative strategies to circumvent this problem.

Our benchmarks also give an overview of how effective the algorithms are given a
certain amount of empty space and specific spatial arrangements. We already
noted that empty space inside nodes is particularly hard to find for the
construction algorithms. The Menger Sponge data set e.g.\ suffers from this
problem and one can see that hardly any of the algorithms is effective at
skipping empty space in this situation.

While the hybrid grid algorithm will usually not outperform the deep
\textit{k}-d tree construction algorithms, its construction rate is in general
superior to the latter. Hybrid grids---for larger data sets---enable frame rates
that fall in-between those of shallow and deep \textit{k}-d trees. For that
reason, we consider them an interesting data structure. As the shallow
\textit{k}-d tree construction algorithm is in most cases almost as fast as the
LBVH algorithm, it might be interesting to combine this construction algorithm
with the hybrid grid algorithm instead of using \textit{k}-d tree construction
on the CPU.

\section{Conclusion}
In this paper, we thoroughly compared the various algorithms to construct
spatial indices for empty space skipping and structured volumes we proposed in
our prior work. We proved that the construction algorithms that make a more
informed decision than just splitting in the middle along one access result in
superior space skipping hierarchies. The paper may serve as a guideline for
pracitioners to decide which algorithm to choose depending on the specific
problem. A general limitation of the top-down construction algorithms is that
the greedy heuristic may accept local optima, which can be partially mitigated
by enforcing a maximum leaf node size. We however showed that this strategy
might fail in certain situations and consider an alternative solution to that
problem interesting future work.

\section*{Acknowledgements}
We thank Ingo Wald for helpful discussion and in particular for pointing out the
trick to enforce a maximum leaf node size by incorporating a middle split.

\footnotesize
\bibliographystyle{eg-alpha}
%\bibliographystyle{ACM-Reference-Format}
%\bibliography{sample-bibliography}
\bibliography{egbibsample}

\clearpage
\begin{table*}[htb]
    %rendering performance for na\"ive ray marching without space skipping. In parentheses we report
    %the rendering times for a transfer function that does not assign empty space at all. Data structures that depend
    %on \emph{min-max} range queries incur an extra constant overhead of $0.042$ seconds to build an IRT for a
    %transfer function array with $1024$ entries.}
	\centering%
    \footnotesize
    \setlength\tabcolsep{3.5pt}
  \begin{tabu}{l|c|c|c|c|c|c|c|c|c|c|c|c|c|c|c|c|c|}
    \toprule

Data Set &
Naive &
\multicolumn{2}{c|}{LBVH} &
\multicolumn{2}{c|}{CPU Shallow} &
\multicolumn{2}{c|}{CPU, MLS=32} &
\multicolumn{2}{c|}{CPU, MLS=128} &
\multicolumn{2}{c|}{GPU Shallow} &
\multicolumn{2}{c|}{GPU, MLS=32} &
\multicolumn{2}{c|}{GPU, MLS=128} &
\multicolumn{2}{c|}{Hybrid} \\

\midrule
~ &
fps &
build & fps &
build & fps & 
build & fps &
build & fps & 
build & fps &
build & fps & 
build & fps &
build & fps \\

\midrule

Aneurism        & 163. 
        & 0.001 & 156.
        & 0.031 & 222.
        & 0.153 & 263.
        & 0.132 & 256.
        & 0.003 & 222.
        & 0.442 & 263.
        & 0.004 & 263.
        & 0.031 & 172.\\

Xmas Tree       & 97.1
        & 0.004 & 69.
        & 0.191 & 103.
        & 0.689 & 112.
        & 0.450 & 116.
        & 0.009 & 98.0
        & 2.163 & 115.
        & 0.009 & 116.
        & 0.191 & 86.2\\

Magnetic        & 105.
        & 0.004 & 103.
        & 0.182 & 192.
        & 0.504 & 158.
        & 0.326 & 185.
        & 0.003 & 192.
        & 1.758 & 156.
        & 0.006 & 182.
        & 0.182 & 141.\\

Beetle          & 82.7
        & 0.009 & 164.
        & 0.423 & 144.
        & 0.910 & 208.
        & 0.667 & 227.
        & 0.030 & 135.
        & 1.985 & 200.
        & 0.004 & 233.
        & 0.423 & 196.\\

Snake           & 65.4
        & 0.012 & 133.
        & 0.884 & 102.
        & 1.987 & 212.
        & 1.938 & 196.
        & 0.043 & 89.3
        & 2.533 & 213.
        & 0.004 & 227.
        & 0.884 & 145.\\

Sponge          & 55.6
        & 0.029 & 23.0
        & 1.116 & 55.6
        & 8.480 & 42.6
        & 3.713 & 66.2
        & 0.050 & 55.6
        & 25.07 & 37.9
        & 0.018 & 55.2
        & 1.116 & 42.6\\

Richtmeyr       & 12.9
        & 0.252 & 6.73
        & 35.82 & 19.9
        & 95.14 & 10.4
        & 72.91 & 17.0
        & 0.396 & 20.0
        & 413.9 & 7.83
        & 29.80 & 15.0
        & 35.82 & 14.3\\

N-Body 265      & 166.
        & 0.001 & 204.
        & 0.004 & 278.
        & 0.069 & 286.
        & 0.059 & 294.
        & 0.002 & 270.
        & 0.142 & 303.
        & 0.562 & 303.
        & 0.004 & 227.\\

N-Body 512      & 100.
        & 0.003 & 144.
        & 0.133 & 185.
        & 0.331 & 227.
        & 0.276 & 238.
        & 0.008 & 175.
        & 1.168 & 233.
        & 0.562 & 233.
        & 0.133 & 164.\\

N-Body 1K       & 55.0
        & 0.016 & 122.
        & 1.426 & 118.
        & 2.627 & 217.
        & 2.279 & 200.
        & 0.054 & 111.
        & 3.637 & 233.
        & 1.959 & 189.
        & 1.426 & 130.\\

N-Body 2K       & 38.8
        & 0.104 & 110.
        & 17.07 & 51.5
        & 24.26 & 213.
        & 22.53 & 204.
        & 0.427 & 84.0
        & 12.28 & 233.
        & 10.90 & 175.
        & 17.07 & 98.0\\
%23  & 7  & 8,895	& 29& 3,571   & 22   \\
%Xmas Tree       & 85,089    & 19 &  33& 11	  & 21,041  & 38 &	13,453	& 37 &
%23  & 7  & 35,677	& 38& 16,753  & 31  \\  
%Magnetic        & 110,253   & 18 &  51& 14	  & 12,329  & 37 &	4,265	& 34 &
%29  & 6  & 20,663	& 31& 7,261   & 26  \\
%Stag Beetle     & 106,923   & 21 &  19& 7	  & 14,229  & 31 &	8,615	& 29 &
%21  & 7  & 23,559	& 36& 13,747  & 28  \\
%Kingsnake       & 92,047    & 22 &  31& 14	  & 33,035  & 46 &	30,359	& 46 &
%17  & 7  & 40,619	& 39& 34,087  & 37  \\
%Menger Sponge   & 2,002,859 & 22 &  1 & 1	  & 115,335 & 22 &	1,399	& 13 & 1
%& 1  & 181389	& 41& 7,269   & 28  \\
%Richtmeyr       & 18,191,777& 25 &  3 & 2	  & 764,809 & 40 &	99,169	& 35 & 1   & 1  & 2,287,401	& 48& 166,745 & 36  \\
%N-Body $256^3$  & 8,753     & 16 &  21& 8	  & 2,917   & 26 &	2,095	& 24 &
%15  & 8  & 5,021	& 30& 3,039   & 22  \\
%N-Body $512^3$  & 41,699    & 19 &  21& 8	  & 13,427  & 34 &	9,367	& 34 &
%17  & 9  & 22,823	& 35& 10,687  & 31  \\
%N-Body $1024^3$ & 127,572   & 22 &  25& 10	  & 39,287  & 59 &	28,991	& 53 &
%17  & 9  & 58,189	& 41& 30,161  & 36  \\
%N-Body $2048^3$ & 361,633   & 25 &  23& 10	  & 66,539  & 72 &	64,047	& 71 &
%19  & 10 & 94,485	& 42& 80,943  & 40  \\

    \bottomrule
  \end{tabu}
  \caption{\label{tab:results}
    Construction and rendering performance for the various data structures.
    Construction performance is reported in seconds and rendering performance is
    reported in frames per second (fps).
    }
\end{table*}

\end{document}